# Deciphering the role of LiBr as redox mediator in Li-O$_2$ Aprotic Batteries

*Angelica Petrongari [a], Lucrezia Desiderio [a], Adriano Pierini [a], Enrico Bodo [a], Mauro Giustini [a], Sergio Brutti [a,b,c]*

[a] Department of Chemistry, Sapienza University of Rome, P.le Aldo Moro 5, Rome, 00185, Italy.

[b] CNR-ISC, Consiglio Nazionale delle Ricerche, Istituto dei Sistemi Complessi, Rome, 00185, Italy.

[c] GISEL – Centro di Riferimento Nazionale per i Sistemi di Accumulo Elettrochimico di Energia, Florence, 50121, Italy.

ABSTRACT

Lithium-oxygen batteries are among the most promising energy storage systems due to their high theoretical energy density, but their practical implementation is hindered by poor reversibility and parasitic reactions. Redox mediators such as LiBr have emerged as a strategy to enhance reaction kinetics and reduce overpotentials. In this study, we explore the impact of three different solvents, dimethoxyethane (DME), tetraethylene glycol dimethyl ether (TEGDME), and dimethyl sulfoxide (DMSO), on the electrochemical performance and reaction pathways of LiBr-mediated Li-$O_2$ cells. Our results reveal that a $^1O_2$ evolution channel that leads to singlet oxygen-induced cell degradation is active only in the TEGDME-based electrolyte. Both DME and DMSO allow singlet oxygen-free Oxygen Evolution Reaction, but only DME is found chemically stable in the LiBr-mediated Li-$O_2$ cell working conditions. These findings highlight the critical role of solvent-mediator interactions in determining the performance of Li-$O_2$ cells.

INTRODUCTION

Aprotic Li-$O_2$ batteries are considered a promising technology to address the increasing demand of energy storage systems. These devices can potentially deliver outstanding performances thanks to their gravimetric energy density (up to 3458 Wh/kg with $Li_2O_2$ as discharge product) and high operating potential (i.e. 2.96 V vs. $Li^+/Li^0$)[1]. The redox processes involved in the functioning of a Li-$O_2$ cell are the dissolution (discharge) and deposition (charge) of lithium at the lithium metal anode; the Oxygen Reduction Reaction (ORR, discharge) and Oxygen Evolution Reaction (OER, charge) at the carbonaceous cathode. In aprotic electrolytes, ORR and OER lead to the formation and dissolution, respectively, of the non-conductive solid lithium peroxide[2].

Some key drawbacks must be faced before achieving the practical implementation of these systems: in particular, the high overpotentials required during the charge process, i.e. the oxidation of lithium peroxide to molecular oxygen, must be efficiently reduced to avoid undesired processes that lead to early performance decay of the cell[3,4]. Many soluble catalysts named Redox Mediators (RMs) are currently being explored and proposed in literature to favour the oxidation of $Li_2O_2$ and preserve the cell components from degradation[5,6]. RMs are oxidized at the cathode surface and then diffuse to $Li_2O_2$ deposits where they oxidize the peroxide anion to molecular oxygen, acting as electron carriers and allowing the lowering of the charge potential[7].

Adding lithium halides, LiX, to the electrolyte solution is a promising and cost-effective strategy to employ the redox couples of halide species, in particular $X^-/X_3^-$ and $X_3^-/X_2$, as redox mediators[8]. LiI is a widely studied RM that showed interesting results[9,10], despite implying also additional concerns for the stability of cell components[11]. More recently, LiBr was also proposed as electrolyte salt and redox mediator by Kwak et al.[12], that highlighted some advantages of this RM against LiI: a higher oxidizing power due to the higher standard potential of the redox couple $Br^-/Br_3^-$; a more efficient formation of $Li_2O_2$ during discharge and little to none competition between bromide chemistry and ORR/OER. In Kwak et al.'s report, LiBr performance was particularly promising when using diglyme as electrolyte solvent, while a decay of the cell performance along with a worsening of the charge/discharge overpotentials was observed using tetraglyme as solvent[12]. This behavior deviates from that observed in the case of LiI[9], and the fundamental origin of such a different performance outcome when varying between very similar solvents was not further explained up to now. Basing on Pierini et al.'s calculations[13], we hypothesized that, similarly to LiI case, the promotion of singlet oxygen release due to the introduction of the RM could be involved. The oxidation of a cluster of $(Li_2O_2)_4$ by $Br_3^-$ resulted thermodynamically favored in both highly polarity solvents like DMSO and ethereal solvents[13], suggesting that in the case of LiBr singlet oxygen generation could be favored also in low-polarity media, in contrast with the case of LiI.

In fact, in our previous study[14], we observed that the use of LiI coupled with a high-polarity solvent like DMSO, where the oxidation of $Li_2O_2$ by $I_3^-$ corresponds to a $\Delta G<0$[13], leads a significative increase of the amount of $^1O_2$ evolved during the oxidation of $Li_2O_2$ by the redox mediator.

The implications of using LiBr on the degradative processes in the Li-$O_2$ cell need to be widely explored in order to design a proper electrolyte formulation to maximize the performance of this promising RM. With this aim, we carried out a study employing LiBr 200 mM + LiTFSI 1M in monoglyme (DME)/tetraglyme (TEGDME)/dimethylsulfoxide (DMSO) solvents, to find out the fundamental origin of different performance outcomes within the same class of solvents and same range of polarity in the case of DME and TEGDME ($\varepsilon_r$ = 7.2 and 7.4, respectively[15,16]), and to highlight also the effect of a sharp change in polarity using DMSO ($\varepsilon_r$ = 46.8[17]).

MATERIALS AND METHODS

*Electrolytes.* High-purity 1,2-dimethoxyethane [anhydrous, 99.5%, inhibitor-free], tetraglyme [tetraethylene glycol dimethyl ether, anhydrous, ≥99%] and dimethyl sulfoxide (Sigma-Aldrich, anhydrous, ≥99%) were purchased from Sigma-Aldrich and dried with 4 Å molecular sieves for at least 1 week before use. Battery grade LiTFSI (lithium bis(trifluoromethanesulfonyl)imide extra dry <20 ppm of $H_2O$, Solvionic) was used as received. LiBr (Reagent-Plus®, ≥ 99%, Sigma-Aldrich) was dried under vacuum at 50°C for 48h before use. Electrolyte solutions of LiTFSI 1M + LiBr 200 mM in DME/TEGDME/DMSO were prepared in an Ar filled glovebox (Iteco Eng SGS-30, $H_2O$ < 0.1 ppm).

*Electrochemical Measurements.* An EL-CELL ECC-Air test cell was used to perform electrochemical experiments. The internal configuration of the cell employed for all the measurements is the following: Li(-)/Separator-Electrolyte/GDL(+)/Ni Foam/Gaseous $O_2$ (1 bar). A glass fiber separator (Whatman, 1.55 mm thickness, 18 mm diameter) soaked in 1 M LiTFSI + 200 mM LiBr in DME/TEGDME/DMSO electrolytes was employed. 15 mm discs of a commercial carbonaceous Gas Diffusion Layer (GDL, MTI Corp.) were used as cathodes. A metallic lithium foil was used as anode. A nickel foam disc (16 mm diameter) was used above the GDL to ensure a homogeneous $O_2$ impregnation. Cell assembly was performed in an Ar filled glovebox (Iteco Eng SGS-30, $H_2O$ < 0.1 ppm). The Li-$O_2$ cells were filled with pure $O_2$, setting a static final pressure of 2.0 bar in the cell volume (head space 4.3 $cm^3$). Galvanostatic cycling tests were run on (-)$Li^0$|LiTFSI 1 M + LiBr 200 mM in DME|GDL(+), (-)$Li^0$|LiTFSI 1 M + LiBr 200 mM in TEGDME|GDL(+) and (-)$Li^0$|LiTFSI 1 M + LiBr 200 mM in DMSO|GDL(+) cells at 0.1 mA $cm^{-2}$ with a limited capacity of 0.2 mAh $cm^{-2}$ and cut-off potentials of 2.0 and 3.6 V vs $Li^+/Li^0$, using a Maccor Series 4000 Battery Test System.

*Ex-situ experiments.* The fluorescent probe DMA [9,10-dimethylanthracene, 99%] and Bromine ($Br_2$, Suprapur®, 99.9999%) were purchased from Sigma-Aldrich and used as received. A set of reference DMA solutions, at the concentrations of 0.5 – 1 – 2.5 – 5 – 7.5 and 10 μM, was prepared in each solvent (DME/TEGDME/DMSO) for the construction of the calibration lines. The reaction solutions were prepared in each solvent according the following scheme:

- <u>Solution A:</u> 5.15 mg of 9,10-dimethylanthracene were dissolved in 5 mL of solvent (DME/TEGDME/DMSO) to obtain a 5 mM solution. The solution was then diluted to 0.2 mM of DMA.

- <u>Solution B:</u> 10 μL of pure $Br_2$ and 33.7 mg of dry LiBr were added to 4.85 mL of solvent to obtain a 40 mM solution of $Br_3^-$. The solution was then diluted to 0.2 mM.

Equal volumes of solutions A and B were mixed to obtain a 0.1 mM solution of both $Br_3^-$ and DMA. A portion of the as-prepared solution was further diluted 10x and its fluorescence and UV spectra were measured as reference before adding $Li_2O_2$. Excess $Li_2O_2$ was added to the other portion of the A+B solution to perform $Li_2O_2$ oxidation by $Br_3^-$ in presence of a 1:1 quantity of DMA as singlet oxygen trap, then the final solution was diluted 10x and its fluorescence and UV spectra were recorded. The preparation of both the reference solutions and the reaction solutions was carried out under Ar atmosphere in a MBRAUN UniLab glovebox [$H_2O$, $O_2 \leq 1$ ppm]. The measurements were carried out in sealable cuvettes Hellma 117104F-10-40, to prevent the exposure of the solutions to air moisture and oxygen. Fluorescence experiments were performed in a Fluoromax-2 Jobin Yvon-Spex spectrofluorometer, with an excitation wavelength of 388 nm. UV spectra were recorded with a Varian Cary 5E UV-Vis spectrometer. Sealable fluorescence quartz cuvettes were employed also for the acquisition of Raman spectra. Spectra of the 40 mM solution of $Br_3^-$ before and after adding $Li_2O_2$ were recorded in TEGDME and DMSO as solvents. Raman measurements were carried out with a DILOR LabRam confocal micro-Raman equipped with a He-Ne laser source at 632.7 nm.

*Computational details.* Simulations of Raman spectra of $Br_3^-$ and $Br_2$ were carried out using density functional theory (DFT) at M062X/def2-TZVP level. Simulations of FTIR spectra of dimethyl sulfide and methanesulfonic acid were carried out by DFT at wB97X/def2-TZVPP[18], NoFrozenCore level of theory, with the Orca package distribution, version 5.03[19].

RESULTS

*LiBr redox mediation in DME.* The charge/discharge capacities achieved during the galvanostatic cycling of a Li-O$_2$ cell containing LiTFSI 1M + LiBr 200 mM in DME, with a limited capacity of 0.2 mAh/cm$^2$ at J = 0.1 mA/cm$^2$ are shown in Figure 1a. This electrolyte formulation allows a stable functioning of the LiBr-mediated cell, with an excellent reversibility of ORR/OER for at least 60 cycles. In line with this, the potential curves reported in Figure 1b indicate both a good discharge profile and a low-overpotential, stable charge performance. Only a mild worsening of the discharge process is observed after 50 cycles, indicated by the higher overpotential at which the 50$^{th}$ discharge takes place. The slight shift of the discharge plateau towards higher capacities observed in the 10$^{th}$ and 50$^{th}$ discharges is likely owed to a moderate accumulation of Br$_3^-$ during charge, which is by consequence reduced in correspondence to the very early stages of discharge at higher potentials than 2.75 V (i.e. the ORR onset). It is important to underline that in the case of the use of LiI as redox mediator, the impact of the I3- accumulation is large[20], while it is almost negligible in the case of LiBr mediation. This aspect indicates that LiBr offers better performance in ethereal solvents, thanks to the very high overpotentials required for the reduction of Br$_3^-$ on the carbon cathode that make this process non-competitive with the ORR during discharge. On passing we would like to stress that, despite the technological validation of this electrolyte formulation for Li-O$_2$ necessarily requires a much more extended cycling in terms of limiting current/capacities and cycle number, here our goal is to demonstrate the fundamental behavior of LiBr as redox mediator in three different solvent media. In this respect the presented galvanostatic tests, even if limited to 60 cycles, offer a solid experimental basis to prove in which chemical environments, LiBr drives or limits parasitic degradation reactivities. This experimental validation aims at consolidating our hypothesis that redox mediators and electrolytes are strongly interplayed and the impact of solvation on the thermodynamics, and kinetics, of redox mediation is not banal and cannot be fully explained by simple considerations starting from the acceptor/donor numbers of solvents.

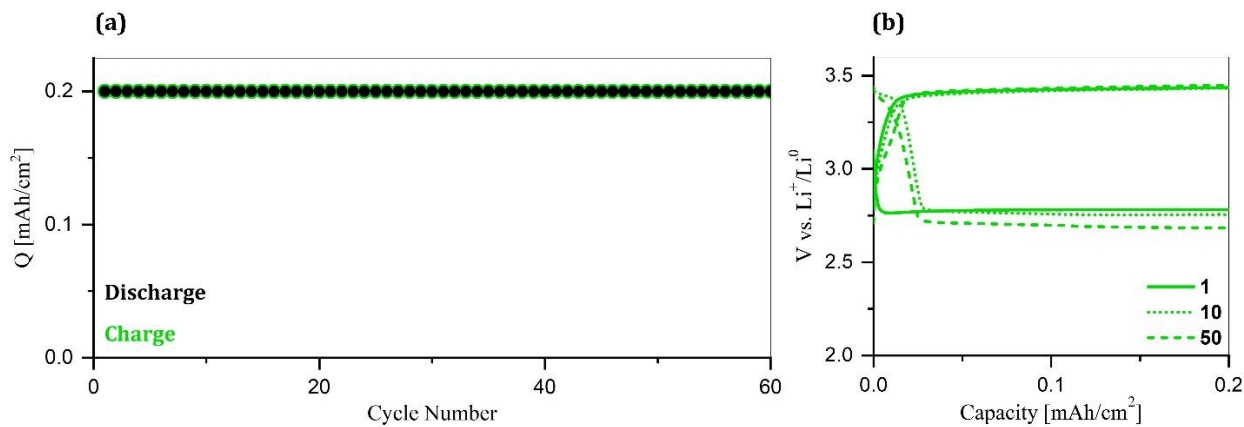

**Figure 1**. (a) Discharge/Charge capacities achieved by a Li-$O_2$ cell with LiTFSI 1M + LiBr 200 mM in DME electrolyte, cycled at J=0.1 mA/cm$^2$, $Q_{lim}$=0.2 mAh/cm$^2$, $V_{cut-off}$=2.0-3.6 V vs. Li$^+$/Li$^0$, for 60 cycles. (b) Voltage profiles of the 1$^{st}$, 10$^{th}$ and 50$^{th}$ cycle.

The good stability of the electrochemical performance of LiBr in DME suggests that this electrolyte formulation leads to limited parasitic chemistries and the possible suppression of $^1O_2$ release beyond the quantities expected by the Boltzmann distribution, as calculated in Eq. S1. To confirm this hypothesis, an ex-situ study on the oxidation of $Li_2O_2$ by $Br_3^-$ in DME was performed. Excess $Li_2O_2$ was added to a solution of 0.1 mM of $Br_3^-$ in DME, in presence of 0.1 mM of the $^1O_2$ trap 9,10-dimethylanthracene (DMA). The reaction solution was then diluted 10x to 0.01 mM and its UV and fluorescence spectra were measured. Simultaneously, another portion of the initial solution was diluted to 0.01 mM (of $Br_3^-$ and DMA) and its UV and fluorescence spectra were measured as a reference of the spectrum of the solution before adding $Li_2O_2$. Figure 2a illustrates the UV spectra of the reaction solution before adding $Li_2O_2$ and after 1h and 1 week. The baseline of pure DME is reported as blank. The UV signal of $Br_3^-$ is localized at ~ 280 nm: this position is assigned from the comparison with the shape and position of $Br_3^-$ UV signal in aqueous solution observed in literature [21]. After 1 hour from adding $Li_2O_2$, a sharp decrease of the $Br_3^-$ signal can be observed, hinting its consumption in the oxidation reaction with lithium peroxide. The complete disappearance of the $Br_3^-$-related signal is observed after 1 week from the onset of the reaction. The UV spectral response during the ongoing of the $Li_2O_2$ oxidation by $Br_3^-$ in DME highlights that not only is the reaction spontaneous but also occurs with relatively fast kinetics. It is worth mentioning that in the case of LiI used as RM, the oxidation of $Li_2O_2$ in glymes is either not spontaneous[13] or strongly kinetically hindered[10], while the analogous reaction results thermodynamically favored in the case of LiBr from both theoretical calculations[13] and experimental evidence. Additionally, the $Li_2O_2$ oxidation by $Br_3^-$ in DME occurs without the evolution of $^1O_2$, as indicated by the comparison of DMA fluorescence spectra before and after adding $Li_2O_2$ (Figure 2b). The DMA signal increases in this case, likely due to the change in the chemical environment after the occurrence of the reaction, promoting the aggregation of DMA molecules. In fact, this effect is expected in the case of stacking of DMA molecules[22].

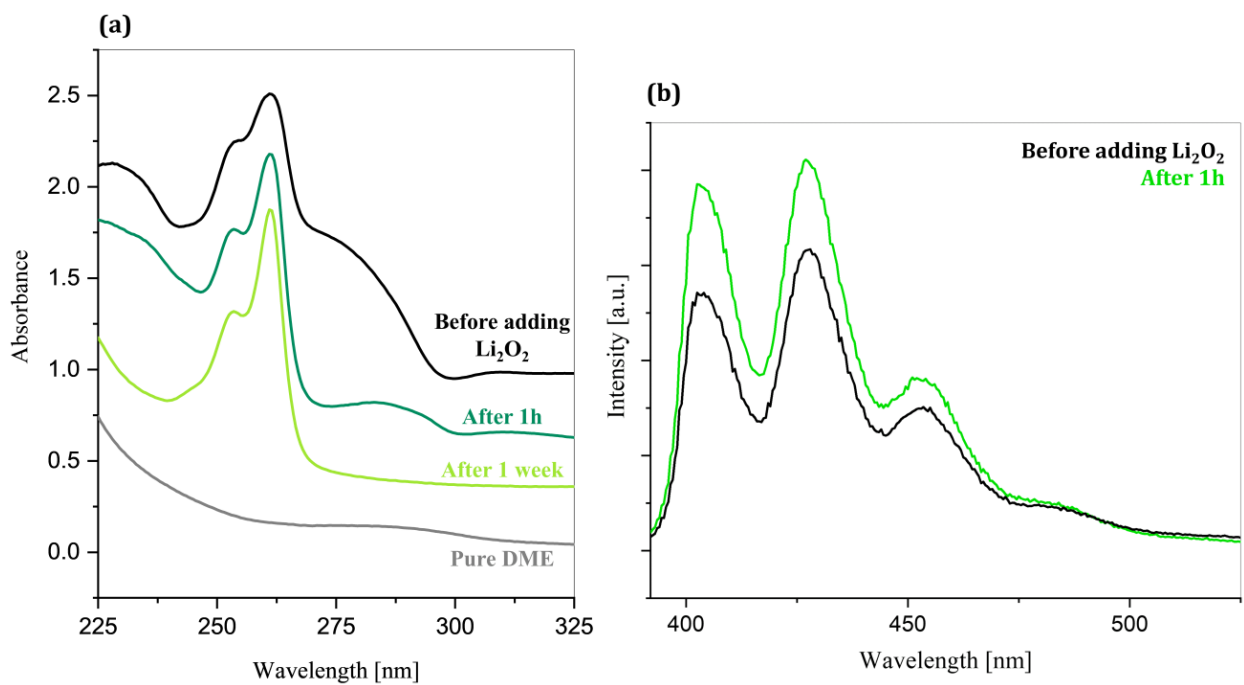

**Figure 2.** (a) UV spectra of pure DME (grey), and a solution of $Br_2$ 10 μM + LiBr 40 μM +DMA 10μM in DME before (black) and after (1h, dark green and 1week, light green) adding excess $Li_2O_2$. (b) Fluorescence spectra of a solution of $Br_2$ 10 μM + LiBr 40 μM + DMA 10μM in DME before (black) and after (green) adding excess $Li_2O_2$.

*LiBr redox mediation in TEGDME.* LiBr performance as redox mediator in TEGDME was firstly studied by Kwak et al.[12], that already observed the worsening of the cell operation in this solvent respect to the shorter-chain diglyme. The observations discussed in the previous paragraph for DME as battery solvent are in line with the picture that emerges from Kwak et al.'s work. Overall, it appears that both DME and diglyme (DEGDME), which differ for a -O-$(CH_2)_2$- unit only, are suitable solvents for LiBr, while earlier performance decay is to be expected for longer chain glymes as TEGDME. This marked performance difference within the same class of solvents is peculiar of LiBr and was not observed in the case of LiI, nor for many other RMs[23]. In Figure 3, the electrochemical behavior of a Li-$O_2$ cell containing LiTFSI 1M + LiBr 200 mM in TEGDME as electrolyte, cycled at J = 0.1 mA/cm$^2$ $Q_{lim}$ = 0.2 mAh/cm$^2$ between 2.0 and 3.6 V vs. Li$^+$/Li$^0$, is reported. Early failure of the charge process is observed starting from cycle 12, from which the charge capacities achieved decrease continuously (Figure 3a). The charge/discharge potential profiles shown in Figure 3b highlight the occurrence of critical parasitic chemistries: at the 10$^{th}$ cycle the participation of Br$^-$ redox to the discharge capacity is extended to almost half of the total capacity, indicating that accumulation of Br$_3^-$ during charge due to inefficient redox mediation is occurring. At the 50$^{th}$ cycle, a three-plateau discharge profile is observed. The first plateau at ~ 3.4 V, which is related to excess Br$_3^-$ reduction, is less extended than at the 10$^{th}$ discharge, however the ORR-related plateau at ~ 2.7 V is shorter due to the presence of a third plateau at ~ 2.4 V, hinting that additional undesired processes are occurring alongside ORR. Observing the charge profiles, it appears that at the 10$^{th}$ charge only an increase in overpotential occurred, while a radical change in the electrochemical process is evident at the 50$^{th}$ cycle.

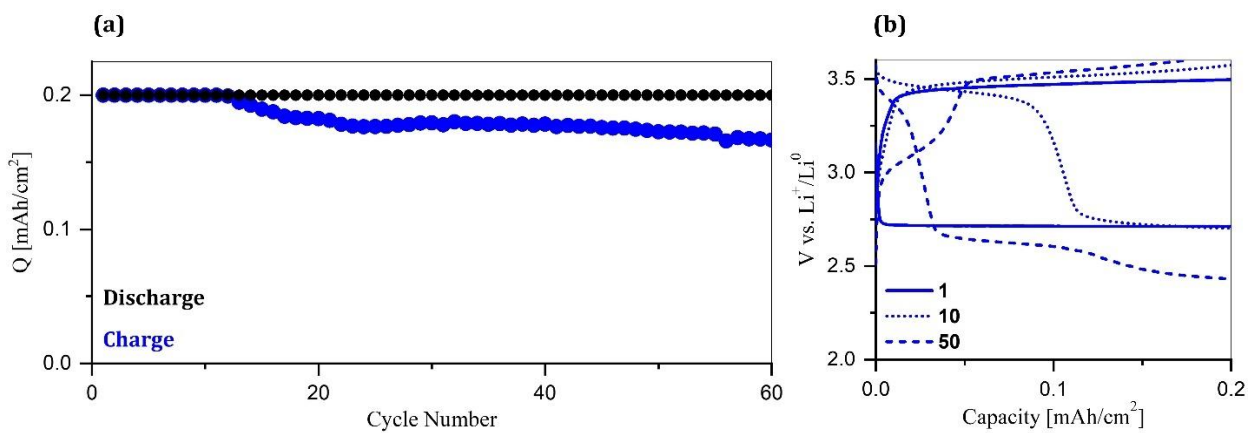

**Figure 3.** (a) Discharge/Charge capacities achieved by a Li-O$_2$ cell with the LiTFSI 1M + LiBr 200 mM in TEGDME electrolyte, cycled at J=0.1 mA/cm$^2$, Q$_{lim}$=0.2 mAh/cm$^2$, V$_{cut-off}$ = 2.0-3.6 V vs. Li$^+$/Li$^0$ for 60 cycles. (b) Voltage profiles of the 1$^{st}$, 10$^{th}$ and 50$^{th}$ cycle.

Such differences between the performance outcomes in DME and TEGDME may indicate that, despite the identical class of ethereal solvents, the $^1$O$_2$ evolution channel activates in a solvent, i.e. TEGDME, that has a slightly larger dielectric constant and a slightly smaller acceptor number. The ex-situ experiments performed to understand the fundamentals of LiBr mediation in TEGDME are reported in Figure 4. To confirm the consumption of $Br_3^-$ in the reaction solution, i.e. 40 mM $Br_3^-$ + 40 mM DMA + excess $Li_2O_2$ in TEGDME, its Raman signal was monitored for 12 hours (Figure 4a): a continuous decrease of its intensity is observed, indicating the occurrence of $Li_2O_2$ oxidation. The assignment of the peak at 175 cm$^{-1}$ to $LiBr_3$ is confirmed by both literature references[8] and theoretical calculations (Figure S1). In Figure 4b, the fluorescence spectra of the singlet oxygen trap DMA before and after adding $Li_2O_2$ in the reaction solution of $Br_3^-$ 0.1 mM + DMA 0.1 mM in TEGDME are reported. A decrease of the fluorescence intensity indicates that part of the DMA in solution reacted with $^1O_2$ forming the non-fluorescent DMA-endoperoxide.

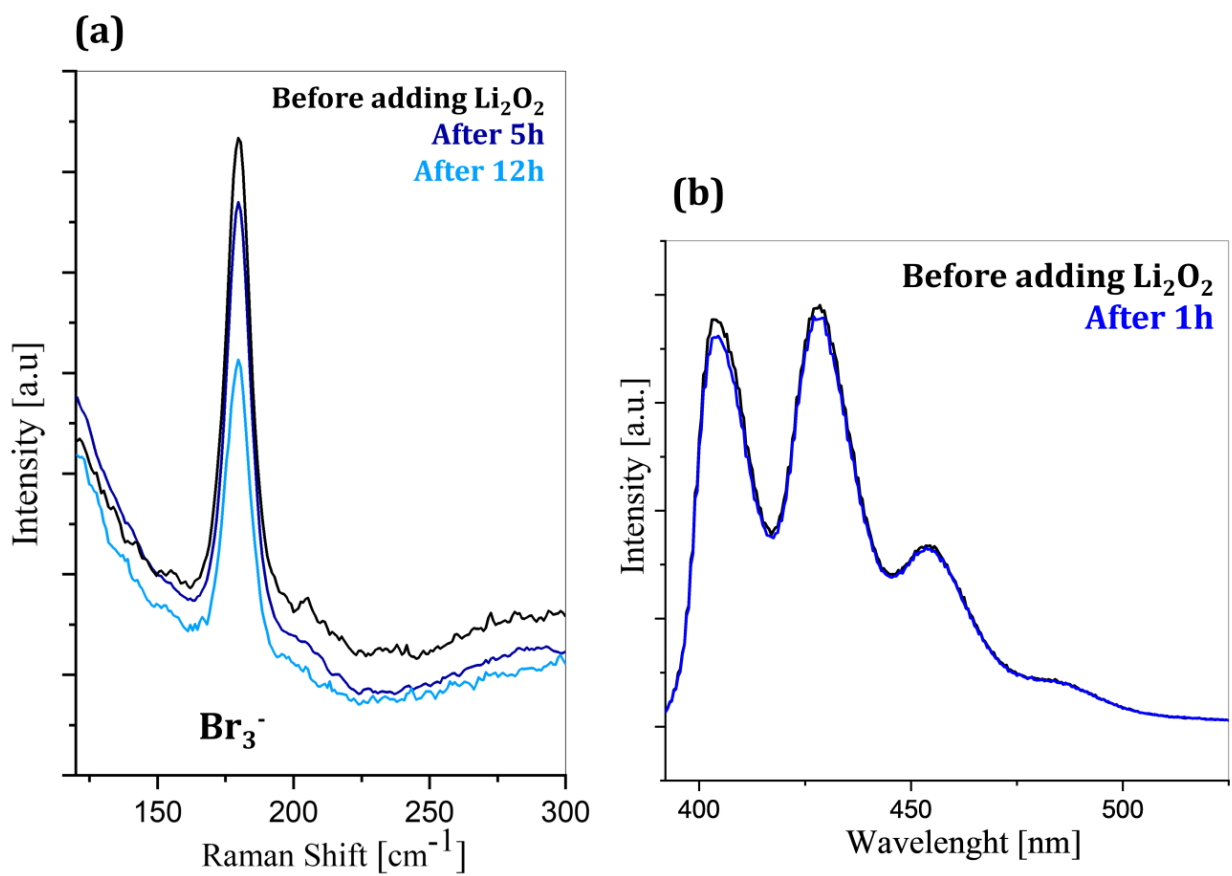

**Figure 4**. (a) Raman spectra of a solution of $Br_3^-$ 40 mM in TEGDME before adding excess $Li_2O_2$, after 5 hours (dark blue) and after 12 hours (light blue). (b) Fluorescence spectra of DMA before (black) and after 1 hour (blue) of adding $Li_2O_2$ to the reaction solution of 0.1 mM $Br_3^-$ + 0.1 mM DMA.

From the calibration line reported in Figure S2, it is calculated that the amount of DMA that reacted with $^1O_2$ is ~ 1.9% of the initial concentration. Since the reaction of DMA with $^1O_2$ follows a 1:1 stoichiometry, this implies that at least ~ 1.9% of the total molecular oxygen evolved in its singlet state. This result is comparable with that obtained for LiI redox mediation in DMSO in our previous study[14] and demonstrates that also LiBr is able to open an alternative pathway for $^1O_2$ evolution enabling its production far beyond the quantities expected from the Boltzmann distribution. Apparently, the key difference lies in the interplay between polarity (dielectric constant) and the Lewis acidity (acceptor number) of the solvent and the Lewis basicity of the redox mediator. LiI is the precursor of the strong Lewis base $I_3^-$ and it leads to $^1O_2$ generation only in a high-polarity solvent, DMSO, that is a strong Lewis acid with large acceptor number. In a low polarity/weak Lewis acid solvent like TEGDME, LiI mediates a $^1O_2$-free $Li_2O_2$ oxidation. On the contrary LiBr is the precursor of the weak Lewis base $Br_3^-$, and activates the release of singlet oxygen in a low-polarity solvent, TEGDME, that is a weak Lewis acid, whereas in DME, a low polarity solvent but with an intermediate Lewis acidity (i.e. acceptor number intermediate between TEGDME and DMSO) the $Li_2O_2$ oxidation mediated by LiBr is $^1O_2$-free.

*LiBr redox mediation in DMSO.* A different behavior of LiBr-mediated Li-$O_2$ batteries is observed in the case of using DMSO as electrolyte solvent. In Figure 5 the cycling performance of a Li-$O_2$ cell with LiTFSI 1M + LiBr 200 mM in DMSO is reported.

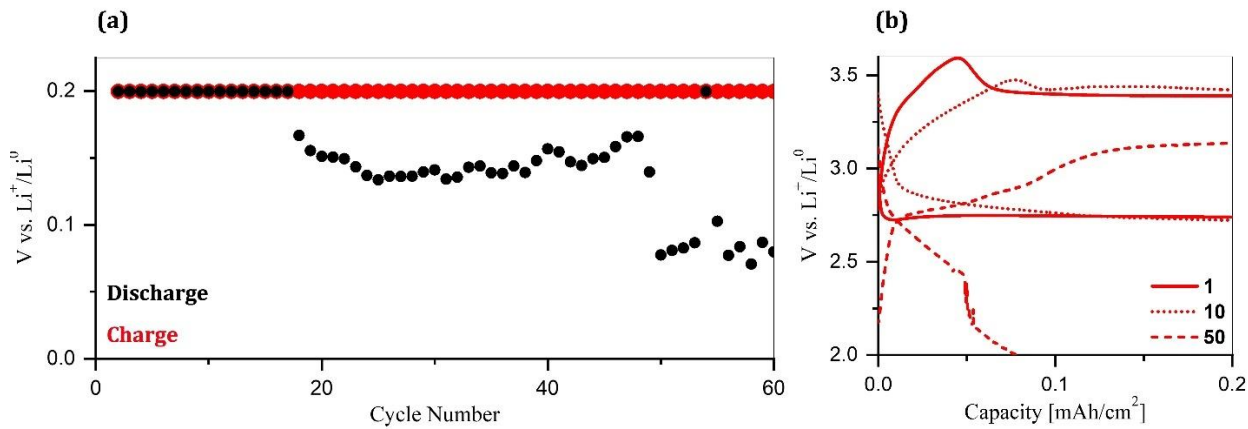

**Figure 5.** (a) Discharge/Charge capacities achieved by a Li-O$_2$ cell with the LiTFSI 1M + LiBr 200 mM in DMSO electrolyte, cycled at J=0.1 mA/cm$^2$, Q$_{lim}$=0.2 mAh/cm$^2$, V$_{cut-off}$=2.0-3.6 V vs. Li$^+$/Li$^0$ for 60 cycles. (b) Voltage profiles of the 1$^{st}$, 10$^{th}$ and 50$^{th}$ cycle.

The discharge process is inefficient starting from the 18$^{th}$ cycle (Figure 5a), likely due to the lack of molecular oxygen in the cell head compartment. The scarcity of oxygen supply may be owed to the inefficiency of the charge process: the shape of the charge potential profiles of the first cycles (Figure 5b) indicate the onset of a parasitic process that is activated at ~ 3.6 V. Based on the observation of the preliminary electrochemical tests with no limit capacity (Figure S3), it appears that this process involves not only the oxidations of Br$^-$ to Br$_3^-$ and Li$_2$O$_2$ to O$_2$, but also the oxidation of electrolyte components, hindering the actual OER. Therefore, the accumulation of Br$_3^-$ (or Br$_2$) occurs cycle by cycle due to the progressive lack of Li$_2$O$_2$ formed during discharge. We can speculate that the high polarity of DMSO allows the involvement of the second redox couple of LiBr within the potential window employed for the galvanostatic cycling of the cell, i.e. 2.0-3.6 V vs. Li$^+$/Li$^0$, with the oxidation of Br$_3^-$ to Br$_2$ occurring at ~ 3.6 V. In fact, it was observed that this reaction occurs at lower potential in DMSO (3.9 V) than in ethers (4.1 V)[7] in Ar atmosphere, and since these values are further lower in presence of O$_2$ [12] it is reasonable that the Br$_3^-$/Br$_2$ oxidation is the process occurring at ~ 3.6 V in the charge profiles reported in Figure 5b. Qualitatively, it was observed that, upon the recovering and washing of the cathodes, only the one cycled in DMSO led to the coloring of the fresh solvent to orange (Figure S4), supporting the presence of high quantities of $Br_2$ in the electrolyte after the cycling procedure. Once formed, $Br_2$ is able to decompose DMSO to dimethyl sulfide, formaldehyde and methanesulfonic acid, according to the scheme elucidated by Aida et al.[24].

Based on these considerations, it is likely that the short cycle life of the Li-$O_2$ cell cycled with the DMSO-based electrolyte is owed to the chemical instability of DMSO in presence of $Br_2$. Although the production of $^1O_2$ is in this case reasonable due to the theoretical predictions[13], the oxidizing power of $Br_2$ towards the solvent and the Li salt may also be the most relevant aspect leading to cell failure.

The ex-situ experiments performed to study the oxidation of $Li_2O_2$ by $Br_3^-$ in DMSO are reported in Figure 6.

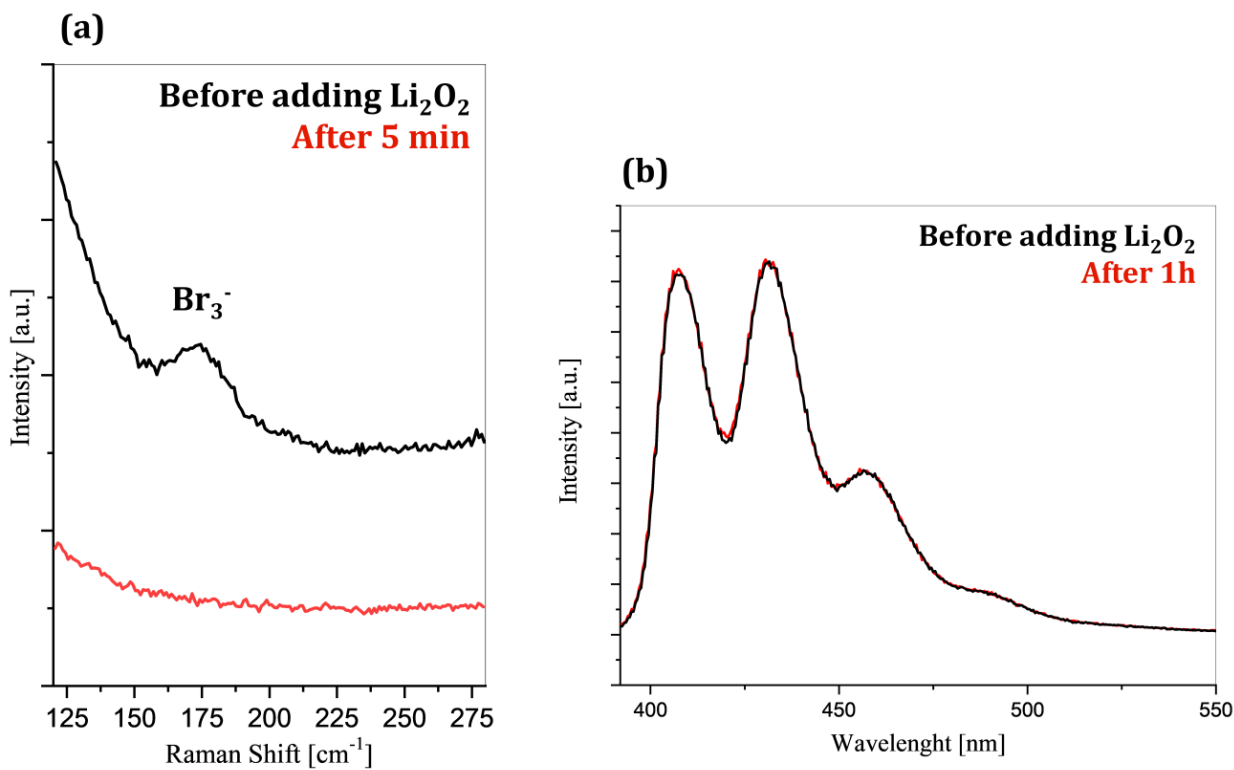

**Figure 6.** (a) Raman spectra of a solution of $Br_3^-$ 40 mM in DMSO before (black) and after 5 minutes (red) from adding excess $Li_2O_2$. (b) Fluorescence spectra of DMA before (black) and after 1 hour (red) from adding $Li_2O_2$ to the reaction solution of 0.1 mM $Br_3^-$ + 0.1 mM DMA.

The Raman signal of $Br_3^-$ 0.4 mM in DMSO was observed before and after adding $Li_2O_2$ (Figure 6a). Due to the band shape, which is strongly different to that observed in ethers (Figure 4a) and has the typical broad aspect of the $Br_2$ signal (Figure S1), it's likely that the $Br_3^-$ species exists mainly as a $Br_2+Br^-$ complex. The consumption of $Br_3^-$ was complete after only few minutes of reaction, indicating that the $Li_2O_2$ oxidation by $Br_3^-$ (or $Br_2$) in DMSO occurs with very fast kinetics. The fluorescence spectra of DMA 0.1 mM in the reaction solution of $Br_3^-$ 0.1 mM before and after adding $Li_2O_2$ are in Figure 6b and clearly indicate that the $Li_2O_2$ oxidation is not accompanied by singlet oxygen evolution in this case. In fact, the fluorescence intensity of DMA is not affected by the occurrence of the reaction. This result is in perfect agreement with our previous considerations about the apparent need for a match between the Lewis character of the RM and the electrolyte solvent despite its polarity: DMSO is a strong Lewis acid characterized by a large acceptor number whereas $Br_3^-$ is a weak Lewis base. Overall, LiBr redox mediation in DMSO, a high polarity solvent, seems to follow an alternative mechanism compared to low polarity ethers, that allows $^1O_2$-free $Li_2O_2$ oxidation. However, in the case of DMSO the performance outcome of LiBr as RM remains unsuccessful owing to the solvent instability and the formation of not only $Br_3^-$, but also the more oxidizing $Br_2$ during charge.

**Table 1.** EDX elemental quantitative analysis of Gas Diffusion Layers cycled with different electrolyte formulations.

| Sample | Atomic % | | | | |
|---|---|---|---|---|---|
| | C | O | S | F | Ni |
| GDL$_{(DME)}$ | 45.29 | 32.42 | 3.78 | 13.11 | 3.32 |
| GDL$_{(TEGDME)}$ | 15.84 | 51.62 | 1.07 | 24.07 | 4.67 |
| GDL$_{(DMSO)}$ | 24.72 | 53.19 | 13.10 | 6.48 | 1.64 |

*Post-mortem characterization of cathodes.* Figure 7 reports the SEM images and corresponding EDX mapping of the Gas Diffusion Layer after their cycling in the DME, TEGDME and DMSO based electrolyte formulations, that will be referred to as GDL$_{(DME)}$, GDL$_{(TEGDME)}$ and GDL$_{(DMSO)}$, respectively.

In line with the electrochemical observations, GDL$_{(DME)}$ (Figure 7a and 7b) maintains a good morphology after cycling, like that of the pristine GDL (Figure S5), along with a clean surface free of significant deposits of degradation products. In fact, the EDX elemental mapping reported in Figure 7b and the corresponding quantitative analysis (Table 1) indicate carbon as the main component, with small amounts of fluorine and sulfur which source is likely a moderate degradation of the TFSI$^-$ anion. GDL$_{(DME)}$ composition deviates less than the other postmortem samples respect to the composition of the pristine GDL (Table S1). The main difference lies in the oxygen content, which increases significantly after cycling compatibly to the formation of a natural Cathode-Electrolyte Interphase (CEI) formed from DME-derivates. This picture is further supported by the ATR-FTIR spectrum of GDL$_{(DME)}$ (Figure 4.5), in which the main spectral features belong to TFSI$^-$ and Li$_2$CO$_3$, except from the band at ~ 1000 cm-1 that is ascribed to the -Si-O-Si- bonds of the glass fiber separator and other minor components in the -C-O-C- region[25] belonging to ether-derived byproducts on the CEI.

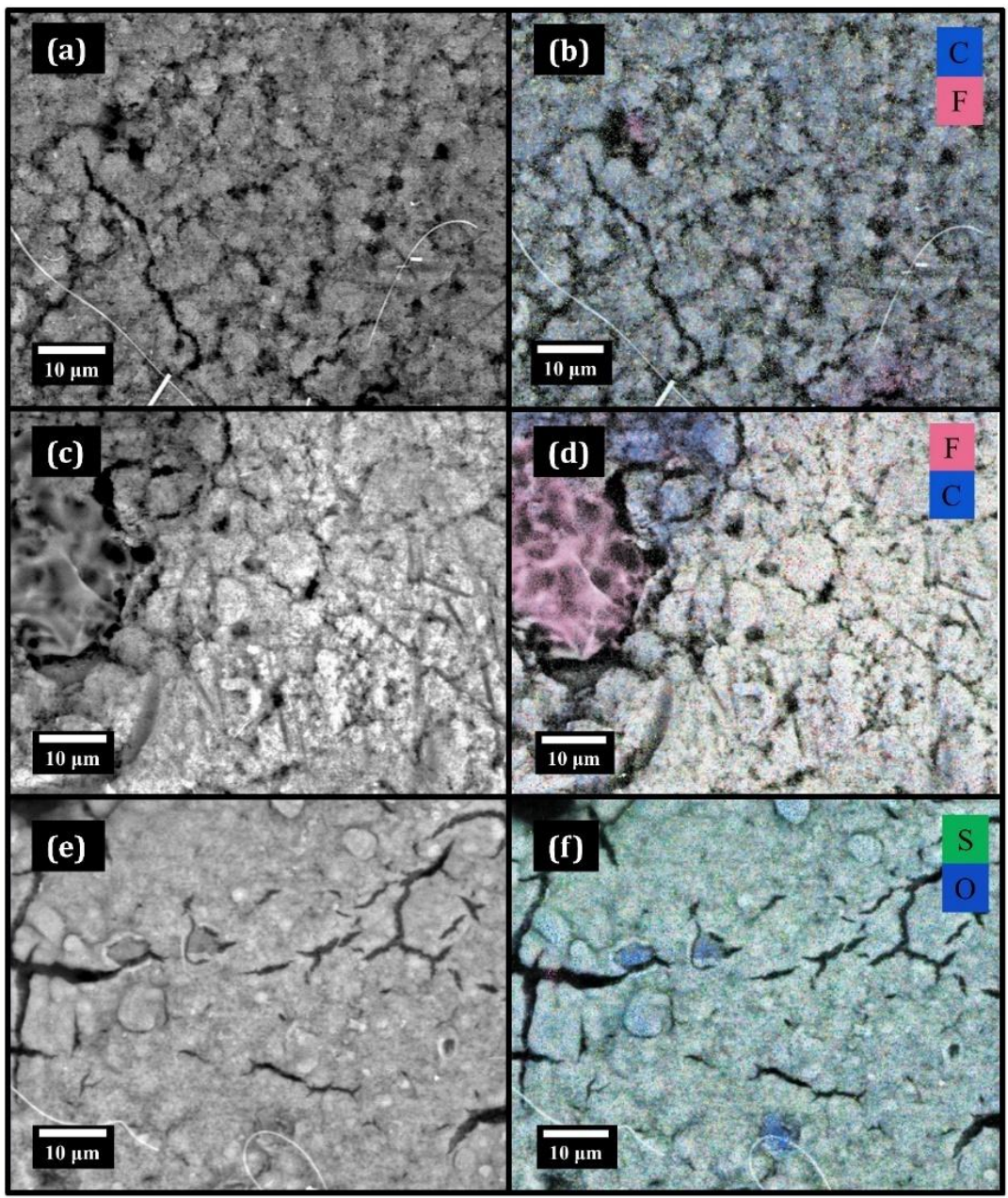

**Figure 7**. Scanning Electron Microscopy images (5.000x magnification, working distance of 8.44 mm, 15kV electron beam energy) of (a) GDL$_{(DME)}$, (c) GDL$_{(TEGDME)}$, (e) GDL$_{(DMSO)}$. Corresponding EDX elemental mapping of (b) GDL$_{(DME)}$, (d) GDL$_{(TEGDME)}$, (f) GDL$_{(DMSO)}$.

Moving to the GDL$_{(TEGDME)}$, in Figure 7c it is possible to observe that large portions of this cathode after cycling are constituted by amorphous and likely organic material, which appeared also very sensitive to the electron beam of the SEM microscope. From the EDX mapping (Figure 7d) and elemental quantification (Table 1) it was found that these regions are fluorine-rich, with no remarkable sulfur content at the same time. This hints that these byproducts are derived from the cathode polymeric binder, which is poly-vinylidenefluoride (PVDF). In fact, the absence of significant amounts of S suggest that the degradation of LiTFSI, which is the only other source of fluorine except PVDF, is likely not involved at all in the formation of these deposits. The degradation of the polymeric binder leads to the loss of cathodic active material, i.e. carbon, in large parts of the cathode surface. The ATR-FTIR spectrum of $GDL_{(TEGDME)}$ is reported in Figure 8b and shows signals that match to those of PVDF and, to a minor extent, to LiTFSI. The small amounts of sulfur which are related to the LiTFSI-deriving byproducts appear located not in correspondence of the fluorine-rich deposits, but only on the remaining carbon surface (see Figure S6). Overall, from the characterization of the cycled cathode it is observed that the polymeric binder of the GDL is the cathode component that undergoes the majority of the undesired reactions occurring during the battery functioning.

The involvement of $^1O_2$ in the degradation processes occurring on $GLD_{(TEGDME)}$ is likely, since carbon and PVDF are usually stable towards the other reactive species such as superoxide and peroxide, that are unavoidably formed during the battery operation in all the electrolyte formulations. This evidence is also in line with the recent findings of Zor et al.[26], that report no relevant $^1O_2$-related degradation of the electrolyte components since neither TEGDME nor LiTFSI are affected by its presence. This implies that in case of $^1O_2$ evolution the other cell components, including the electrode binder, would be preferentially attacked.

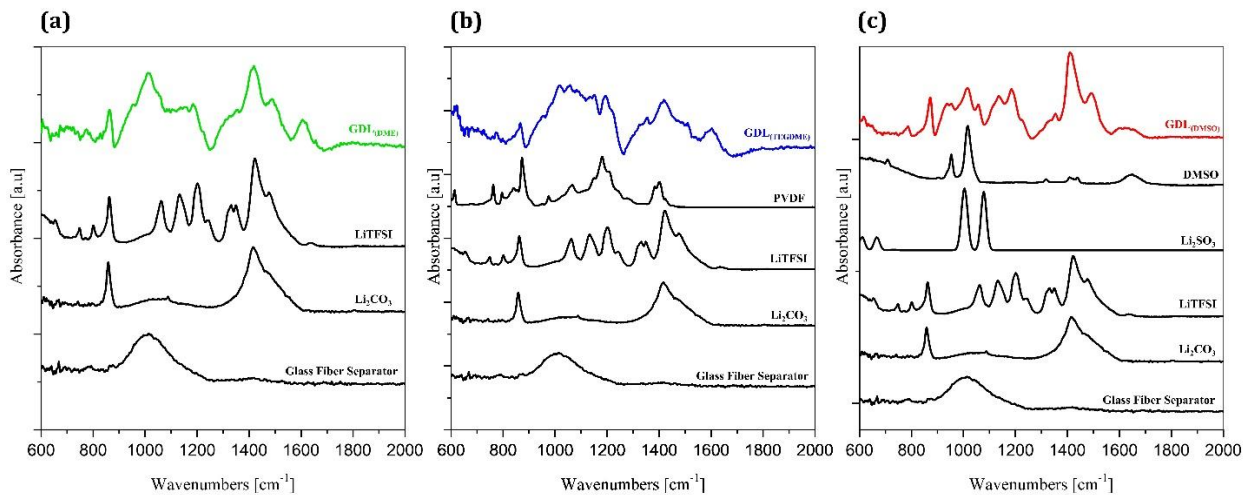

*Figure 8.* ATR-FTIR spectra of (a) GDL$_{(DME)}$ and reference spectra of LiTFSI, Li$_2$CO$_3$ and Whatman Glass Fiber separator. (b) GDL$_{(TEGDME)}$ and reference spectra of PVDF, LiTFSI, Li$_2$CO$_3$ and Whatman Glass Fiber separator. (c) GDL$_{(DMSO)}$ and reference spectra of DMSO, Li$_2$SO$_3$ (theoretical), LiTFSI, Li$_2$CO$_3$ and Whatman Glass Fiber separator.

The SEM images and EDX elemental maps of GDL$_{(DMSO)}$ are reported in Figure 7e and 7f and confirm the accumulation of byproducts during the battery operation. The surface of the GDL is covered by a thick film-like layer formed from the degradation of mainly DMSO and secondarily TFSI$^-$, according to the relative quantities of S and F detected by EDX elemental quantification (Table 1). The Cathode-Electrolyte Interphase is composed by uniformly-distributed sulfur and localized oxygen-rich regions (Figure 7f). In the ATR-FTIR spectrum of GDL$_{(DMSO)}$ (Figure 8c) the diagnostic signal of S=O stretching at 1044 cm$^{-1}$ [115] is intense, indicating the presence of other DMSO-related byproducts. Among them, Li$_2$SO$_3$ is likely also present since there are compatible peaks between 900 and 1100 cm$^{-1}$ in the sample spectrum. The presence of dimethyl sulfide and methanesulfonic acid, the main byproducts expected by the Br$_2$ induced decomposition of DMSO[24], is also compatible with the GDL$_{(DMSO)}$ spectrum as highlighted in Figure S7.

*Post-mortem characterization of anodes.* Figure 9 shows the morphology and elemental mapping by SEM/EDX of Li metal anodes after cycling in the different electrolytes, that will be referred to as Li$_{(DME)}$, Li$_{(TEGDME)}$ and Li$_{(DMSO)}$.

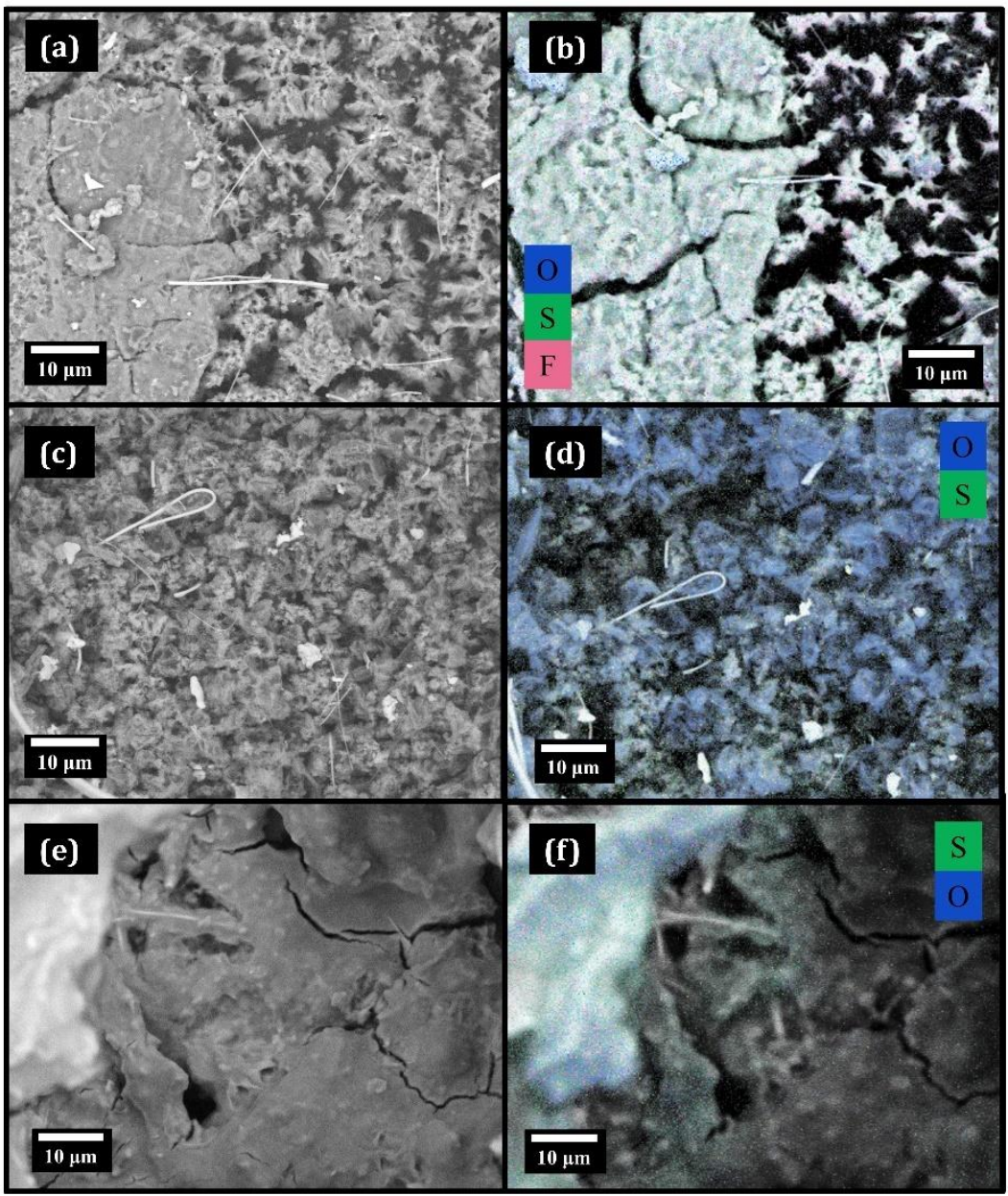

**Figure 9.** Scanning Electron Microscopy images (5.000x magnification, working distance of 8.44 mm, 15kV electron beam energy) of (a) Li$_{(DME)}$, (c) Li$_{(TEGDME)}$, (e) Li$_{(DMSO)}$. Corresponding EDX elemental mapping of (b) Li$_{(DME)}$, (d) Li$_{(TEGDME)}$, (f) Li$_{(DMSO)}$.

**Table 2**. EDX elemental quantitative analysis of lithium metal anodes cycled with different electrolyte formulations.

| Sample | Atomic % | | | | |
|---|---|---|---|---|---|
| | C | O | S | F | Ni |
| Li$_{(DME)}$ | 36.09 | 16.58 | 10.67 | 32.10 | 3.49 |
| Li$_{(TEGDME)}$ | 12.75 | 76.36 | 2.79 | 6.94 | -- |
| Li$_{(DMSO)}$ | 46.17 | 20.20 | 17.75 | 11.70 | 3.67 |

The SEM image of Li$_{(DME)}$ in Figure 9a clearly shows anode deterioration with the presence of dendritic structures, indicating that the anode functioning is not optimal in the DME-based electrolyte. As in the other Li metal battery systems, in Li-O$_2$ cells dendrites lead to Li detachment and loss of electrical contact that compromise the cell functioning, and also constitute a safety hazard due to the possibility of short circuits[27]. The corresponding EDX elemental mapping in Figure 9b indicates oxygen-rich regions, that may hint that the degradation of the ether solvent occurred. The significant amounts of S and F on the Li surface (see Table 2) indicate also the presence of TFSI$^-$ fragments in the Solid Electrolyte Interphase (SEI). The morphology of Li$_{(TEGDME)}$ in Figure 9c appears different and shows predominantly mossy lithium. Thus, also in this case the uniformity of the Li morphology is compromised after the battery cycling, but the composition of the SEI layer is remarkably better in respect to the Li$_{(DME)}$ case. In fact, both the elemental mapping in Figure 9d and the quantitative analysis in Table 2 show oxygen as the major component. Usually, the presence of oxygen corresponds to that of lithium, that cannot be detected by EDX spectroscopy. Thus, a rich and relatively uniform mapping of oxygen on the Li metal anode suggests the presence of a very thin SEI layer originating from byproducts formed by the reactivity of Li with the electrolyte. Few amounts of S and F on the Li surface indicate mild TFSI$^-$ degradation that always occurs during the formation of the natural SEI layer on Li anodes. The improved stability of the TEGDME-based electrolyte towards Li metal respect to DME case is explained by TEGDME tendency to complex Li$^+$ within the LiTFSI ionic couple thanks to its longer chain[28]. This intermolecular interaction was demonstrated to play a key role in protecting the ether from oxidation[29], i.e. the hydrogen abstraction from methylene groups by nucleophilic agents[30].

The Li$_{(DMSO)}$ anode in Figure 9e shows some separator fibers that are incorporated in a thick film, composed by S and O mainly (Figure 9f). Significant quantities of carbon are also present in the electrode surface (Table 2), supporting the occurrence of severe degradation of electrolyte components. As in the case of GDL$_{(DMSO)}$, the relative amount of S is higher than that of F, indicating that sulfur-containing byproducts derive mostly from DMSO. The Li metal anode appears consumed in the parasitic processes and covered with an insulating film that affects the reversibility of the plating/stripping processes.

CONCLUSION

Employing LiBr as redox mediators in Li-O$_2$ batteries leads to radically different outcomes varying the solvent media. The optimal reversibility of cathodic reactions, i.e. ORR and OER, is obtained in DME, while critical issues rise in both the other solvents, i.e. TEGDME and DMSO. The cathode functioning in TEGDME is affected by $^1$O$_2$-induced parasitic processes.

Conversely, in DMSO LiBr redox mediation appears to follow a different mechanism allowing $^1O_2$-free OER but, at the same time, enabling DMSO decomposition induced by $Br_2$. The Li metal anode functioning is not directly correlated to the cathode operation, since the worst Li morphology was found in the DME case. This implies that Li metal protection must be addressed independently.

In summary, this work demonstrates that the release of $^1O_2$ during the Li-$O_2$ cells cycling results from a complex interplay or molecular properties of the constituents in the Li-$O_2$ electrolytes: apparently this parasitic chemistry emerges in the case of a matching between the Lewis character of the solvent and the redox mediator, despite the polar character of the solvent media. Therefore, the most promising approach for a reversible operation of an aprotic Li-O2 battery is the employment of RM with Lewis basicity contrasting the corresponding Lewis acidity of the electrolyte solvent, in parallel with a tailored Li metal protection strategy the limits the parasitic chemistries originating at the negative side.

ACKNOWLEDGEMENTS

This research work has been supported by the University of Rome La Sapienza through the grants RM12117A5D5980FB, RM122181677EDA1D and AR1221816B9459F2 and by the "Progetto Orangees - ORgANics for Green Electrochemical Energy Storage – codice CSEAA_00010" funded by the Italian Government through the MITE (Ministero della Transizione Ecologica) call 2022 "Bandi di gara di tipo A". EB and A.Pierini also acknowledge the financial support of "La Sapienza" with grants n. RM12117A33BCD47C and AR12117A8AE9CAA3.

## ASSOCIATED CONTENT

The following files are available free of charge.

Supporting Information (PDF)

## AUTHOR INFORMATION

Corresponding Author

*Email: sergio.brutti@uniroma1.it

## ACKNOWLEDGMENTS